\documentclass[fleqn]{aa520}
\usepackage{epsf}
\usepackage{txfonts}
\usepackage{graphicx}
\newcommand{\kms}{km\,s$^{-1}$}
\newcommand{\erg}{erg\,s$^{-1}$}
\newcommand{\ergs}{erg\,cm$^{-2}$s$^{-1}$}

\newcommand{\hatoms}{H-atoms\,cm$^{-2}$}

\newcommand{\lsun}{$L_{\odot}$}

\newcommand{\msun}{$M_{\odot}$}

\newcommand{\mk}{$M_{\rm K}$}

\newcommand{\rsun}{$R_{\odot}$}

\newcommand{\smyr}{M$_{\odot}$yr$^{-1}$}

\begin{document}
\title{A precise HST parallax of the cataclysmic variable EX~Hydrae,
its system parameters, and accretion rate\thanks{Based on observations
made with the NASA/ESA Hubble Space Telescope, which is operated by
the Association of Universities for Research in Astronomy, Inc., under
NASA contract NAS 5-26555. These observations are associated with
proposal \#9230.}}

\author {K.\ Beuermann \inst{1}
     \and   Th. E. Harrison  \inst{2,**}
     \and   B. E. McArthur \inst{3}
     \and   G. F. Benedict \inst{3}
     \and   B.T.\ G\"ansicke \inst{4}}

\institute{Universit\"ats-Sternwarte G\"ottingen, Geismarlandstr.~11, 
D-37083 G\"ottingen, Germany, beuermann@uni-sw.gwdg.de
\and
New Mexico State University, Box 30001/MSC 4500, Las Cruces, NM 88003,
tharriso@nmsu.edu\thanks{Visiting Astronomer, Cerro Tololo
Inter-American Observatory, National Optical Astronomy Observatory,
which is operated by the Association of Universities for Research in
Astronomy, Inc., under cooperative agreement with the National Science
Foundation.}  
\and 
McDonald Observatory, University of Texas, Austin, TX 78712,
mca@barney.as.utexas.edu, fritz@astro.as. utexas.edu
\and
Department of Physics and Astronomy, University of Southampton,
Highfield, Southampton SO17\,1BJ, UK, btg@astro.soton.ac.uk}

\date{Received March 26, 2003 / Accepted September 18, 2003}
 
\authorrunning{K. Beuermann et al.}
\titlerunning{An HST/FGS parallax of EX Hydrae}

\offprints {beuermann@uni-sw.gwdg.de}

\abstract{Using the HST Fine Guidance Sensor, we have measured a high
precision astrometric parallax of the cataclysmic variable EX~Hydrae,
$\pi=15.50\pm0.29$\,mas. From the wavelength-integrated
accretion-induced energy flux, we derive a quiescent accretion
luminosity for EX Hya of $L_{\rm acc} = (2.6\pm 0.6)\times
10^{32}\,$\,\erg. The quiescent accretion rate then is $\dot M_{\rm
av}=(6.2\pm 1.5)\times 10^{-11}(M_1/0.5\,{\rm M}_\odot)^{-1.61}\, {\rm
M}_{\odot}{\rm yr}^{-1}$. The time-averaged accretion rate, which
includes a small correction for the rare outbursts, is 6\% higher. We
discuss the system parameters of EX Hya and deduce $M_1 =
0.4-0.7$\,\msun, $M_2 = 0.07-0.10$\,\msun, and $i =
76.0^\circ-77.6^\circ$, using recent radial velocity measurements of
both components and restrictions imposed by other observational and
theoretical constraints. We conclude that the secondary is
undermassive, overluminous, and expanded over a ZAMS star of the same
mass. Near the upper limit to $M_1$, the accretion rate of the white
dwarf coincides with that due to near-equilibrium angular momentum
loss by gravitational radiation and angular momentum transfer from the
orbit into the spin-up of the white dwarf. Near the lower mass limit,
the correspondingly higher accretion rate requires that either an
additional angular momentum loss process is acting besides
gravitational radiation or that accretion occurs on a near-adiabatic
time scale. The latter possibility would imply that EX Hya is in a
transient phase of high mass transfer and the associated spin-up of
the white dwarf.
 } 

\maketitle

\section{Introduction}
\label{sec-intro}

Cataclysmic variables (CVs) are ideal sites for the study of accretion
processes.  Understanding these phenomena requires the knowledge of
distances and luminosities. Unfortunately, trigonometric parallaxes,
the only secure source for distances, are so far available only for a
handful of CVs (Harrison et al. 1999, 2000, 2003, McArthur et
al. 1999, 2001).
EX Hya is one of the best-studied intermediate polars, i.e. CVs in
which the rotation of the magnetic white dwarf is not synchronized
with the orbital period. Accretion spins up the white dwarf on a time
scale of $(3.8\pm 0.2)\times 10^6$\,yrs (Hellier \& Sproats 1992).

Earlier distance estimates range from 105\,pc (Warner 1987) to
\mbox{$>130$\,pc} (Berriman et al. 1985), while the recent detection
of the ellipsoidal light modulation of the secondary star in the
infrared suggests a much smaller distance of 65\,pc (Eisenbart et
al. 2002, henceforth EBRG02). Only a trigonometric distance will allow
an accurate measurement of the luminosity and the accretion rate to be
made. The latter is an essential ingredient for the discussion of the
secular evolution of CVs and of the angular momentum loss processes
driving accretion.  Since EX Hya is not included in ground-based
parallax programs, we have obtained a trigonometric parallax using the
{\it HST Fine Guidance Sensor} (FGS).

\section{Observations and data reduction}

The process for deriving a parallax for a cataclysmic variable from
FGS observations has been fully described in papers by McArthur et
al. (2001, 1999) and Harrison et al. (1999). The process used here is
identical to those efforts. An FGS program consists of a series of
observations of the target of interest, and a set of four or more
reference stars located close to that target. Typically, three epochs
of observations, each comprised of two or more individual pointings
({\it HST} orbits), are used to solve for the variables in the series
of equations that define a parallax solution. For EX Hya we were able
to obtain observations on four different epochs (2000 July, 2000
December, 2001 July, and 2002 January). The extra epoch was fortuitous
in that the reference frame for EX Hya was somewhat ``noisy''.  As
described in McArthur et al. (2001) extensive calibration data, as
well as estimates of the distances and proper motions of the reference
stars, are required to obtain a robust parallax solution.

\subsection{Spectroscopic parallaxes of the reference frame}

We have used a combination of spectroscopy and photometry to estimate
spectroscopic parallaxes for the reference stars. Optical {\it BVRI}
photometry was obtained on 2001 March 13 using the CTIO 0.9 m
telescope and the Cassegrain Focus CCD imager\footnote{Go to
http://www.ctio.noao.edu/cfccd/cfccd.html for further information on
this instrument.}. These data were reduced in the normal fashion, and
calibrated to the standard system using observations of Landolt
standards. The final photometric data set is listed in
Table~1. Included in Table~1 is the Two Micron All-Sky Survey (2MASS)
{\it JHK} photometry of the reference stars, transformed to the
homongenized system of Bessell \& Brett (1988) using the
transformation equations from Carpenter (2001). Typical error bars on
the photometry range from $\pm$ 0.02 mag for the {\it V}-band
measurements, to $\pm$ 0.04 mag for the 2MASS photometry of the
fainter stars (except for EX-193, where the error on {\it B $-$ V} is
$\pm$ 0.05 mag).

Optical spectroscopy of the reference frame stars, and a number of MK
spectral type templates, was obtained on 2001 March 9 and 10 using the
CTIO 1.5-m telescope with the Cassegrain Spectrograph\footnote{For a
complete description of the CTIO 1.5-m Cassegrain Spectrograph go to
http://www.ctio.noao.edu/spectrographs/ 60spec/manual/}. We used the
831 {\it l}/mm ``G-47'' grating (resolution 0.56 \AA/pix) with a two
arcsecond slit. From comparison of the spectra of the program objects
to those of the MK-templates, we estimated the spectral types of the
reference stars listed in Table~1.

By combining the spectral types of the reference frame stars with
their photometry, we can derive the visual extinction to each
target. Using the standard relations from Reike \& Lebofsky (1985),
the final visual extinctions $A_{\rm V}$ were computed (see Table~2).
Except for the most distant of the reference stars, the $A_{\rm V}$
values cluster near 0.20 mags.  The interstellar hydrogen column
density to EX Hya as measured by the Ly$\alpha$ absorption profile is
only $3\times10^{18}$\,\hatoms\ (EBRG02), suggesting a much lower
extinction, A$_{\rm V} < 0.01$.

Using the spectral types and visual extinctions of the reference
stars, we obtain the spectroscopic parallaxes given in the last column
of Table~2. To determine these values, we used the {\it Hipparcos}
calibration of the absolute magnitude for main sequence stars by Houk
et al. (1997), and that for giant stars tabulated by Drilling \&
Landolt (2000).  For the astrometric solution discussed below, we
assumed error bars of $\pm$ 25\% on our spectroscopic parallaxes.

\begin{table*}[t]
\caption{Photometric and spectroscopic data for the EX Hya reference frame.}
\label{tab:input}
\begin{tabular}{lcccccccc}
\hline \noalign{\smallskip}
Star & $V$ & $B$-$V$ & $V$-$R$ & $V$-$I$ & $J$-$H$ & $H$-$K$ & $K$
&Spectral Type\\ 
\noalign{\smallskip} \hline \noalign{\smallskip}
EX Hya & 13.26 & 0.13 & 0.25 & 0.79 & 0.34 & 0.21 & 11.75 & \\
EX-177 & 11.42 & 0.92 & 0.53 & 0.96 & 0.48 & 0.08 &  9.34 & K0.5V \\
EX-193 & 14.35 & 1.19 & 0.65 & 1.18 & 0.58 & 0.14 & 11.75 & K4V \\
EX-201 & 13.50 & 0.70 & 0.39 & 0.75 & 0.37 & 0.05 & 11.88 & G1.5V \\
EX-301 & 12.74 & 1.18 & 0.62 & 1.16 & 0.61 & 0.13 & 10.08 & K0III \\
EX-354 & 12.08 & 1.02 & 0.52 & 1.01 & 0.55 & 0.07 &  9.79 & G8III \\
\noalign{\smallskip} \hline \noalign{\smallskip}
\end{tabular}
\end{table*}

\begin{table*}[t]
\caption{Positions, proper motions, visual extinctions and derived 
spectroscopic parallaxes for the EX Hya reference frame.}
\label{tab:input}
\begin{tabular}{lccccccc}
\hline \noalign{\smallskip}
Star & $\alpha_{\rm 2000}$ & PM$_{\alpha}$ (mas/yr)$^{1}$ & 
$\delta_{\rm 
2000}$ & PM$_{\delta}$ (mas/yr)$^{1}$ & A$_{\rm V}$ & $\pi$ (mas)$^2$ \\
\noalign{\smallskip} \hline \noalign{\smallskip}
EX Hya &12 52 24.4& $-$127.3& $-$29 14 56.7 & +31.6 &\hspace{-3.5mm} 
$<0.01^3$&\\ 
EX-177 &12 52 35.5& $-$93.7 & $-$29 12 59.5 & +11.1 & 0.20 & 8.54 \\ 
EX-193 &12 52 09.6& $-$18.9 & $-$29 13 36.6 & $-$7.8 & 0.21 & 3.94 \\ 
EX-201 &12 52 31.4& $-$13.9 & $-$29 14 01.8 & $-$25.3 & 0.20 & 1.66 \\ 
EX-301 &12 52 39.4&    +1.0 & $-$29 14 58.2 & +5.7 & 0.49& 0.47 \\ 
EX-354 &12 52 50.1& $-$10.2 & $-$29 15 38.9 & $-$1.2 & 0.21& 0.56 \\ 
\noalign{\smallskip} \hline \noalign{\smallskip}
\end{tabular}

$^{1}${Zacharias et al. 2000}. $^{2}${Adopted error of $\pi$ is
25\%}. $^{3}${Based on the Ly\,$\alpha$ absorption measurement of
Eisenbart et al. 2002}
\end{table*}

\subsection{The astrometric solution}

The data reduction process for deriving a parallax from FGS
observations was identical to previous efforts, except for the fact
that the astrometer has changed from FGS3 to FGS1R. FGS1R has been
successfully calibrated (McArthur et al. 2002), and is the astrometric
instrument of choice for current {\it HST} programs. 
 
Solving the astrometric equations for the absolute rather than
relative motion requires knowledge of the proper motions of the
reference stars in addition to their parallaxes. We have used the USNO
CCD Astrograph Catalogue (the ``UCAC'', Zacharias et al. 2000) in its
on-line version (VizieR\footnote{Ochsenbein et al. (2000).
http://vizier.u-strasbg.fr/cgi-bin/VizieR} catalogue I/268) to extract
proper motions with errors for all six of our targets (columns 3 and 5
of Table~2). 

With these input data, the astrometric equations (McArthur et
al. 2001, Eqs. 1--4) were solved simultaneously, using GaussFit
(Jefferys et al. 1987) to minimize the $\chi^{\rm 2}$ values of the
solution. Because of the large difference in $B-V$ between EX~Hya and
the reference stars, the lateral colour correction discussed in
Benedict et al. (1999) is included. The final parallax of EX~Hya is
\mbox{$\pi=15.50\pm0.29$\,mas}. The error of 0.29\,mas exceeds that of
previous parallax measurements with the HST FGS of the CVs TV\,Col and
WZ\,Sge (McArthur et al. 2001, Harrison et al. 2003). The reason for
the somewhat larger error for EX Hya is the lower number of epochs in
our case and the noisy character of the reference frame.

As discussed by Lutz \& Kelker (1973), the nature of parallax
measurements implies that the most probable true parallax is slightly
smaller than the observed parallax of a star. The magnitude of the
effect depends on the relative error of the observed parallax and on
the model parameters describing the magnitude and space density
distributions of the parent population (e.g. Hanson 1979). For the
present observation of EX~Hya, the probable value of the correction,
$-0.04\pm0.02$\,mas, is minute compared with the statistical error of
the measurement of 0.29\,mas and is subsequently neglected. 

\section{Discussion}

The measured parallax corresponds to a distance of
\mbox{$d=64.5\pm1.2$}\,pc and confirms the earlier result of EBRG02
based on the somewhat uncertain interpretation of the ellipsoidal
modulation of the secondary star.  Combined with information on the
masses of primary and secondary star, the now accurately known
distance allows us to derive reliable values of the accretion
luminosity and the accretion rate of EX~Hya.

\subsection{Absolute magnitude of EX~Hya}

The distance modulus of EX~Hya is $m-M=4.05\pm0.04$. Our observed
visual magnitude of EX~Hya in quiescence, $m_{\rm V}=13.26$ (see
Table~1), implies $M_{\rm V}=9.22$. EX~Hya is slightly variable even
in quiescence, however, and Hellier et al. (2000) quotes a long-term
orbital-mean quiescent magnitude $m_{\rm V}=13.0$, suggesting an
average $M_{\rm V}$ in quiescence of 9.0.  The absolute magnitude in
outburst based on a mean peak magnitude of $m_{\rm V} \simeq 9.6$
(Hellier et al. 2000) is $M_{\rm V} \simeq 5.6$. This value has still
to be corrected for inclination effects.

\subsection{Kinematic solution}

Recently, Belle et al. (2003) and Vande Putte et al. (2003) reported
velocity amplitudes of the white dwarf and of the secondary star, $K_1
= 59.6\pm2.6$\,\kms\ and $K_2 = 360\pm35$\,\kms, respectively. The
inclination $i$ is defined by the length of the partial, but
flat-bottomed eclipse of the X-ray emission from the lower magnetic
pole, i.e., the one below the orbital plane (Beuermann \& Osborne
1988, Rosen et al. 1988), with a FWHM width of 157\,s (Mukai et
al. 1998). The kinematic solution yields $M_1=0.49\pm 0.13$\,\msun,
$M_2=0.081\pm0.013$\,\msun, and $i=76.6^\circ \pm 0.8^\circ$, where we
have assumed that the lower pole is just visible through the inner
hole in the accretion disk. For each additional $10^8$\,cm required to
be freely visible below the pole, the inclination decreases by
$0.15^\circ$.

A white dwarf mass near $0.5$\,\msun\ was also advocated by Ezuka \&
Ishida (1997), who obtained $M_1=0.50\pm0.08$\,\msun\ from X-ray line
intensity ratios, and by Cropper et al. (1999), who derived $M_1 =
0.46\pm0.04$\,\msun\ from the X-ray continuum temperature. Compared
with the kinematic solution, however, these indirect methods may be
afflicted by systematic errors which are difficult to
estimate. Presently, we allow, therefore, for the larger error in
$M_1$ as given above.

As shown below, the remaining error in $K_2$ is the principal
uncertainty in the discussion of the system parameters, the accretion
rate, and the evolutionary status of EX~Hya. For completeness, we note
that the previous mass determinations of Hellier (1996) and Vande
Putte et al. (2003) yield the same principal result, but were not
based on the most recent $K_1, K_2$ combination. In particular,
Hellier (1996) used the Smith et al. (1993) $K_2=356\pm 4$\,\kms\,
which has since been superseded by the reanalysis of the same data by
Vande Putte et al. (2003) yielding the larger error quoted above. 

\subsubsection{The primary star}

For $M_1=0.49\pm 0.13$\,\msun, Wood (1995) white dwarf models with a
``thick'' hydrogen envelope and an effective temperature of $10^4$\,K
predict $R_1=(1.02\pm 0.15)\times 10^9$\,cm.  Given the distance, we
can compare this radius with that implied by the HST/GHRS UV flux and
the mean effective temperature $T_{\rm eff, wd} = 25000\pm3000$\,K
(EBRG02, Belle et al. 2003) of the accreting white dwarf. For $d =
64.5$\,pc, we obtain an effective radius $R_{\rm 1,eff} \simeq
6.0\times 10^8$\,cm, which corresponds to $M_1 \simeq
0.95$\,\msun. Hence, either the white dwarf is much more massive than
suggested by the kinematic solution or the UV emission originates only
from the heated polar caps and the underlying white dwarf is much
cooler. Below, we shall exclude $M_1 \ga 0.8$\,\msun\ and conclude,
therefore, that the second possibility must hold. Large warm spots
seem to be the rule in pole-accreting white dwarfs in AM Her stars and
probably exist also in IPs (e.g., G\"ansicke et al. 1995, G\"ansicke
et al. 1998, G\"ansicke 2000). This is consistent with the fact that
the typical white dwarf temperature expected from compressional
heating in short-period CVs is $10000-20000$\,K (Townsley \& Bildsten
2002, G\"ansicke 2000, G\"ansicke et al. 2000, Szkody 2002). Modelling
the HST/GHRS spectrum of EX~Hya with a polar cap of 25000\,K and a
cooler underlying white dwarf of $10^9$\,cm radius at the measured
distance of EX~Hya yields an approximate upper limit to the effective
temperature of the latter of 17000\,K.

\subsubsection{The secondary star}

The low kinematic mass of the secondary star may surprise at first
glance. Such a low mass, however, is in line with the secondary masses
in Z Cha (Wade \& Horne 1988), HT Cas (Horne et al. 1991), and OY Car
(Wood et al. 1989). These results for short-period CVs suggest
that with decreasing orbital period the radii of the Roche-lobe filling
secondaries exceed those of ZAMS objects (Baraffe et al. 1998) and
that the secondaries are slightly out of thermal equilibrium. The
secondary in EX~Hya falls in line with this trend. For
$M_2=0.081\pm0.013$\,\msun, its Roche radius is $R_{\rm 2,R} =
(9.58\pm 0.50)\times 10^9$\,cm\ while a Roche-lobe filling ZAMS-star
would have a radius smaller by $15-23$\% (Baraffe et al. 1998,
Renvoiz\'e et al. 2002).

Given the distance, the radius of the secondary can also be estimated
from its luminosity and effective temperature. At $d=64.5$\,pc, the
$K$-band magnitude of the secondary star $K\simeq 12.4\pm 0.3$
(EBRG02) implies an absolute magnitude $M_{\rm K}=8.4\pm 0.3$. Using
the stellar models with solar metallicity of Baraffe et al. (1998), we
convert \mk\ to a bolometic luminosity $L_2\simeq
0.0024\pm0.0006$\,\lsun. With an effective temperature for the
dM$4\pm1$ secondary in EX~Hya (EBRG02) of $T_{\rm eff} \simeq
3200-3300$\,K (Leggett et al. 1996, 2000), we then obtain a radius
$R_2=(10.6\pm 1.6)\times 10^9$\,cm, consistent with the kinematically
determined Roche-lobe radius.

In summary, the kinematic solution for EX~Hya is consistent with the
picture which evolves from studies of other short-period CVs. While
the spectral type of the secondary in EX~Hya is as expected for a
Roche-lobe filling main sequence star in a CV of the same orbital
period (Beuermann et al. 1998), the kinematic solution suggests that
the secondary is undermassive, overluminous and expanded
over a ZAMS object.

\begin{table}[t]
\caption{Comparion of luminosity derived and theoretically predicted
accretion rates. Column (2) is the effective exponent of the mass
radius relation of the secondary star, colums (2) and (3) are for the
nominal kinematic solution, $M_1,M_2 = $0.49,\,0.081\msun, and colums
(4) and (5) for the $+1\sigma$ limit of $K_2$, $M_1,M_2 =
$\,0.62,\,0.094\msun. Accretion rates are in \smyr, inner disk radii
$r_{\rm i,9}$ are in units of $10^9$\,cm.}
\begin{tabular}{lccrcr}
\hline \noalign{\smallskip}
& (1) & (2) & (3) & (4) & (5)\\
Model  & $\zeta$ & lg\,$\dot M_1$  & $r_{\rm i,9}$ & lg\,$\dot M_1$ &  
$r_{\rm i,9}$ \\[1ex]  
\hline \noalign{\smallskip}
\multicolumn{6}{l}{{\it (a) Luminosity-derived accretion rate}}\\[1ex]
        &      & $-10.17$ && $-10.34$ \\
        &      & \hspace{1.6mm}$\pm 0.10$ && \hspace{1.6mm}$\pm 0.10$ \\[1ex]
\multicolumn{6}{l}{{\it (b) Model accretion rates and inner disk radii}}
\\[1ex]
$r_{\rm i,max}, \dot M_{\rm min}$ &      & $-10.54$ &27.4& $-10.65$ & 30.0 \\
$r_{\rm i,min}, \dot M_{\rm max}$&      & $-10.13$ &4.4& $-10.19$ & 4.0\\[1ex]
gravitational only             & 0.85 & $-10.93$ && $-10.73$ & \\ 
                   & 0.50 & $-10.86$ && $-10.66$ & \\
  & \hspace{-2.5mm}$-0.20$& $-10.65$ && $-10.46$ &\\[1ex]
grav + spin-up     & 0.85 & $-10.45$ &19.6& $-10.43$& 12.2\\
                   & 0.50 & $-10.37$ &13.9& $-10.36$ & 8.6\\
  & \hspace{-2.5mm}$-0.20$& $-10.17$ &5.4& $-10.16$ &  3.4\\[1ex]
$3\times$grav + spin-up&0.85&$-10.23$ &7.2&$-10.13$& 3.0\\
                   & 0.50 & $-10.16$ &5.1& $-10.06$ & 2.2\\
  & \hspace{-2.5mm}$-0.20$& \hspace{0.8mm} $-9.95$ &2.0& $-9.86$ & 0.9\\
\noalign{\smallskip} \hline \noalign{\smallskip}
\end{tabular}
\end{table}

\subsection{Quiescent accretion luminosity}

The mean orbital observed flux of EX~Hya integrated over all
wavelengths from the infrared to the hard X-ray regime is $(6.0\pm
0.6)\times 10^{-10}$\,\ergs\ (EBRG02, their Tab. 4). This value
includes the Fe L excess flux around 1 keV (Mukai 2001), an estimate
of the hard X-ray flux beyond 20\,keV, and an estimate of the (small)
XUV flux shortward of the Lyman edge. It also includes the flux
received from the white dwarf, but excludes that from the secondary
star. According to EBRG02, about $1.8\times 10^{-10}$\,\ergs\ is from
to the white dwarf. As noted above, we attribute much of this flux to
reprocessed radiation from its irradiated pole caps. In order to
obtain the quiescent accretion-induced flux, the fraction originating
from the unheated white dwarf photosphere has to be subtracted. For a
range of effective temperatures of $10000-17000$\,K (see above), an
0.5\,\msun\ white dwarf contributes $(0.7\pm 0.5)\times
10^{-10}$\,\ergs\ to the wavelength-integrated flux. Subtracting this
contribution, we obtain the accretion-induced flux during quiescence,
$f_{\rm q}=(5.3\pm 0.8)\times10^{-10}$\,\ergs. The corresponding
quiescent accretion luminosity of EX~Hya is $L_{\rm q} = 4\pi
d^2f_{\rm q}=(2.6\pm 0.6) \times10^{32}$\,\erg, where we have
quadratically added an additional error of 20\% for basing $L_{\rm q}$
on the mean orbital flux instead of on the $4\pi$-averaged flux.

\subsection{Accretion rate}

\subsubsection{Quiescence}

The observationally determined accretion rate in quiescence, $\dot
M_{\rm q}$, is related to the luminosity by \mbox{$L_{\rm q} \simeq
GM_1\dot M_{\rm q}/R_1$}. We take $M_1$ for the moment as a free
parameter and approximate the Wood (1995) radii of white dwarfs with a
thick hydrogen envelope as $R_1 \simeq 10.0\times 10^8(M_1/0.50\,
{\rm M}_{\odot})^{-0.61}$\,cm, valid for masses between 0.40 and
0.65\,\msun. The luminosity can then be converted to a quiescent
accretion rate $\dot M_{\rm q}=(6.2\pm1.4)\times
10^{-11}(M_1/0.5M_\odot)^{-1.61}$\,\smyr. This value is valid
as an average over the last 30 years over which the quiescent
magnitude of EX~Hya has stayed approximately constant.

\subsubsection{Time-average including dwarf nova outbursts}

$\dot M_{\rm q}$ has still to be corrected for the effect of the rare
dwarf nova outbursts, in which the visual magnitude of EX~Hya rises
from 13\,mag to about 10\,mag for 1--2 days.  Hellier et al. (2000)
assumed that the mass transfer rate scales as the optical brightness and
estimated that the outbursts typically involve $10^{22}$\,g. Using his
gap-corrected outburst recurrence rate of 1.5\,yrs suggests that the
time-averaged accretion rate $\dot M_{\rm 1,av}\simeq 1.05\times \dot
M_{\rm q}=(6.5\pm 1.5)\times 10^{-11}(M_1/0.5\,{\rm
M}_\odot)^{-1.61}\,{\rm M}_{\odot}{\rm yr}^{-1}$.

\begin{figure}[t]
\includegraphics[width=8.8cm]{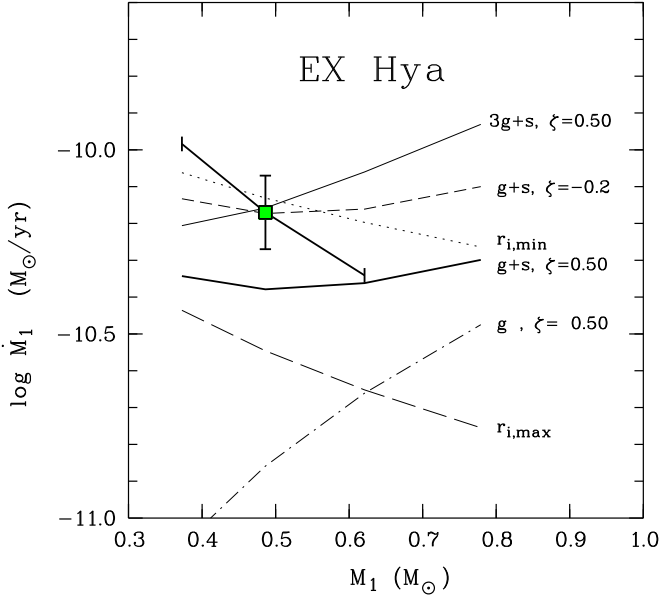}
\caption[ ]{Accretion rate as a function of $M_1$ for the spin-up rate
of Hellier \& Sproats (1992) and different assumptions on the driving
mechanism, gravitational radiation only (g), gravitational radiation
plus spin-up (g+s), and a systemic angular momentum loss of $3 \times$
the gravitational rate plus spin-up (3g+s). Accretion from the
corotation radius defines a mininal accretion rate (long-dashed curve,
$r_{\rm i} = r_{\rm co}$). The luminosity-derived accretion rate of EX
Hya is shown as the solid square with error bars (see text).}
\end{figure}

\subsection{Theoretical accretion rate}

The accretion rate is physically related to the spin-up rate of the white
dwarf which has been observed over a time span of 30 years,
$-P_{\rm 1}/\dot P_{\rm 1} = (3.8\pm 0.2)\times 10^6$\,yrs (Hellier \& Sproats
1992). Equivalently, $\dot P_{\rm 1}=-(3.4\pm 0.2)\times 10^{-11}$ and
$\dot \Omega_{\rm 1}=(1.32\pm 0.07)\times 10^{-17}$\,s$^{-2}$.

The relation between spin-up of the white dwarf and accretion rate has
been extensively discussed by Ritter (1985) for the case of
conservative mass transfer, i.e., $\dot M_1 = -\beta \dot M_2$ with
$\beta = 1$. The alternative extreme of $\beta = 0$ applies to the
case that the accreted mass is shed off again from the white
dwarf. Non-conservative mass transfer was treated by Rappaport et
al. (1983) and considered in recent evolutionary calculations (e.g., Kolb \&
Baraffe 1999, Howell et al. 2001). We adopt the conservative case here
because (i) a wind is unlikely to flow off from the magnetic white
dwarf, (ii) propeller action is unlikely to be very efficient given
the slow rotation of the white dwarf, and (iii) the relation between
accretion and spin-up considered here is that over the last 30 years
and not a secular one, which obviates the need to account for mass and
angular momentum loss in nova eruptions. The difference in the
predicted mass transfer rates for EX~Hya between the $\beta = 1$ and
$\beta = 0$ cases amount to only $\Delta$(log\,$\dot M_2) \simeq
0.050$ because the specific orbital angular momentum of the white
dwarf is small.

\begin{table}[t]
\caption{System parameters of EX Hydrae (see text).}
\label{tab:input}
\begin{tabular}{llr}
\hline \noalign{\smallskip}
Parameter              & Value or range  & \hspace{-3mm}Ref.\\[1ex]
\hline \noalign{\smallskip}

\multicolumn{3}{l}{{\it (a) Observed parameters, literature:}}\\
Orbital period $P_{\rm orb}$     & 5895.4~s & (1) \\
Spin period $P_{\rm 1}$        & 4021.6~s & (1) \\
Spin-up rate $P_{\rm 1}/\dot P_{\rm 1}$ & $(3.8\pm 0.2)\times 10^6$~yrs & (1)\\
Velocity amplitude $K_1$       & $59.6\pm 2.6$~\kms & (2)\\
Velocity amplitude $K_2$       & $360\pm 35$~\kms & (3)\\
Width of X-ray eclipse          & 157~s & (4)\\[1ex]
\multicolumn{3}{l}{{\it (b) Kinematic solution based on literature values:}}\\
White dwarf mass $M_1$         & $0.49\pm 0.13$~\msun & \\
Secondary mass $M_2$           & $0.081\pm 0.013$~\msun & \\
Inclination $i$                & $76.6^\circ\pm 1.0^\circ$ & \\[1ex]
\multicolumn{3}{l}{{\it (c) Observed parameters, this work:}}\\
HST FGS Parallax $\pi$                 & $15.50\pm 0.29$\,mas & \\
Distance $d$                   & $64.5\pm 1.2$\,pc & \\
Accretion rate $\dot M_{\rm 1,av}$& $(6.6\pm 1.5)\times 10^{-11}\times$\\
&\multicolumn{2}{l}{\hspace*{8mm}$(M_1/0.5\,M_\odot)^{-1.61}$\,\smyr}\\[1ex] 
\multicolumn{3}{l}{{\it (d) Comparison with theory, this work:}}\\
White dwarf mass $M_1$         & $0.40-0.70$~\msun & \\
Secondary mass $M_2$           & $0.07-0.10$~\msun & \\
Implied range of $K_2$         & $335-413$~\kms & \\
White dwarf radius $R_1$       & $(0.017-0.012)$~\rsun & \\
Roche radius $R_2$             & $(0.132-0.148)$~\rsun & \\
Inclination $i$                & $76.0^\circ-77.6^\circ$ & \\
Inner edge of disk $r_{\rm i}$ & $(5-9) \times 10^9$~cm & \\
Accretion rate $\dot M_1$      & $(8-4)\times 10^{-11}$\,\smyr 
& \\
\noalign{\smallskip} \hline \noalign{\smallskip}
\end{tabular}

(1) Hellier \& Sproats (1992), (2) Belle et al. (2003), (3) Vande
 Putte et al. (2003), (4) Mukai et al. (1998).
\end{table}

Using Ritter's (1985) Eq.\,(1) or (A19), we calculate the accretion
rate $\dot M_1$ driven by a systemic angular momentum loss rate $\dot
J_{\rm sys}$, for given $M_1, M_2$, $\dot \Omega_{\rm 1}$, and an
effective mass radius exponent of the secondary star,
$\zeta=\,$d\,ln\,$R_{\rm 2}$/d\,ln\,$M_{\rm 2}$ ( with $\zeta \equiv
\alpha_2$ in Ritter 1985). The transfer rate $\dot M_1$ increases with
decreasing $\zeta$. Equilibrium models of low mass main sequence stars
yield $\zeta\simeq 0.85$ (Baraffe et al. 1998). As mentioned above,
however, the observed masses of Z Cha, HT Cas, OY Car, and WZ Sge and
the requirement that the stars fill their Roche lobes suggest some
bloating beyond that caused already by deformation of the star in the
Roche potential (Renvoiz\'e et al. 2002), consistent with $\zeta\sim
0.5$ and a mild loss of thermal equilibrium. The minimum value which
$\zeta$ can assume in fully convective stars is near the adiabatic
limit, $\zeta_{\rm ad} \simeq -1/3$, and may be realized after turn-on
of mass transfer (d'Antona et al. 1988).  An additional model
parameter is the efficiency of the driving mechanism. We chose two
cases: (i) gravitational radiation only, supplemented by the observed
spin-up and (ii) the same plus an unspecified additional mechanism
twice as effective as gravitaional radiation. The resulting $\dot M_1$
for these models are listed in Table~3 for two
$M_1,M_2$-combinations, the nominal kinematic solution and the
$+1\sigma$ value of $K_2$ (Vande Putte et al. 2003), i.e., $M_1 =
0.49$\,\msun\ and $M_1 = 0.62$\,\msun, respectively. For completeness,
we include the ``gravitational only'' case (without spin-up).

\subsection{EX~Hya as an intermediate polar}

Figure\,1 shows the luminosity-derived $M_1$-dependent accretion rate
by the filled square, with the inclined ``horizontal'' error bar
indicating the $\pm 1\sigma$-range of $M_1$. Also shown are the
accretion rates for selected models and for a larger range of $M_1$,
with ``g'' referring to angular momentum loss by gravitational
radiation and and ``s'' to the spin-up contribution. Several
interesting conclusions can be drawn from a comparison of these
rates. While the angular momentum drain by gravitational radiation can
not be avoided, the spin-up observed over 30 years signifies an
additional transfer of orbital angular momentum into the spin of the
white dwarf which increases $\dot M_1$. The combined effect is
described by the g\,+\,s model (Ritter 1985, his Eq.\,1, see also
the details in the Appendix to his paper). Since the secondary in EX
Hya is slighly expanded over a main sequence object, we chose an
effective $\zeta = 0.50$. This case represents an approximate lower
limit to $\dot M_1$ for the current state of EX~Hya (thick solid curve
in Fig.~1). $\dot M_1$ for $\zeta\simeq 0.85$ would be lower by only
16\% ($\Delta$(log\,$\dot M_2) \simeq 0.075$, see Table~3).

Independent information on $\dot M_1$ can be gathered from equating
the time derivative of the white dwarf's spin angular momentum to the
torque exerted by matter which couples to the white dwarf's
magnetosphere at a lever arm $r_{\rm i}$, i.e., $J_{\rm s,1} = \dot
M_1 r_{\rm i} v_{\rm K}(r_{\rm i})$, where $v_{\rm K}(r_{\rm i})$ is
the Keplerian velocity at $r_{\rm i}$ and $\dot J_{\rm s,1}$ depends
on the spin-up rate $\dot \Omega_1$ (and the structural change of the
white dwarf associated with $\dot M_1$) (see Eqs.\,6, 7 and A15 of
Ritter 1985). For given $\dot \Omega_1$ and $\dot M_1$, the equality
defines the radial lever arm $r_{\rm i}$, which we identify with the
inner edge of the accretion disk or, vice versa, chosing limits on the
lever arm defines limiting values for $\dot M_1$. To be sure, this is
strictly correct only if the inner disk is in Keplerian motion and
coupling occurs only at $r_{\rm i}$ (and not also further out), which
may not be the case (e.g., King \& Wynn 1999). We do not further
consider this question here and only note that in these more general
cases the actual inner edge of the circulating material, loosely
referred to as disk here, would be located inside the so-defined
$r_{\rm i}$. Two conditions thus restrict the permitted range of $\dot
M_1$ independent of all other arguments. A lower limit $\dot M_{\rm
min}$ is given by $r_{\rm i,max}=r_{\rm L1}$, the distance to the
L$_1$ point (long dashed curve in Fig.~1). Correspondingly, an upper
limit $\dot M_{\rm max}$ is given by $r_{\rm i,min}=R_1/{\rm cos}\,i$,
the inner disk radius which just allows the lower pole of the white
dwarf to be viewed through the inner hole in the disk (dotted curve in
Fig.~1). This free view is required by the partial, yet flat-bottomed
X-ray eclipse (Mukai et al. 1998, see also Beuermann \& Osborne 1988,
Rosen et al. 1988). If one requires the emission at a distance larger
than $R_1$ below the orbital plane to be visible through the hole,
$r_{\rm i}$ has to be correspondingly larger, which would move the
dotted curve in Fig.~1 further downward.  The permitted range of
$r_{\rm i}$ (see Table~4) is consistent with that deduced
spectroscopically by Hellier et al. (1987).

With the above conditions, the range permitted for $\dot M_1$ in
Fig.~1 is restricted to below the dotted and above the thick solid
curve. Interestingly, this range vanishes for $M_1 \ga 0.8$\,\msun\
and, hence, excludes a white dwarf mass above this
limit. Gratifyingly, $\dot M_{\rm 1,av}$ for the range of $M_1$ as
given by the kinematic solution is entirely consistent with this
permitted area in the $\dot M_1, M_1$-plane. Without knowledge of the
kinematic masses, the permitted area alone restricts $M_1$ to stay
between about 0.40\,\msun\ and 0.70\,\msun. This mass range
corresponds to $K_2=335-415$\,\kms\ in excellent agreement with
Vande Putte's (2003) $K_2=360\pm 35$\,\kms. Table~4 summarizes
our results. The error in $\dot M_1$ could be reduced to that in the
bolometric flux if a precise measurement of $M_1$ or $R_1$ would be
available.

Finally, we comment on the evolutionary state of EX~Hya and the
properties of the mass-losing secondary star.  For $M_1 \simeq
0.6-0.7$\,\msun, the implied $\dot M_{\rm 1,av}\simeq 4\times
10^{-11}$\,\smyr\ is entirely consistent with near equilibrium mass
transfer under the action of gravitational radiation and spin-up
(model g\,+\,s with $\zeta \simeq 0.5-0.9$). For $M_1\simeq
0.4-0.5$\,\msun, near the best-fit kinematic and X-ray-derived primary
masses, the higher transfer rate $\dot M_{\rm 1,av}\simeq 8\times
10^{-11}$\,\smyr\ can be provided in two scenarios: (i) A more
efficient angular momentum loss process is at work, which we describe
as gravitational radiation plus an additional loss process which is
about twice as efficient as the former (model 3g\,+\,s with $\zeta
\simeq 0.5-0.9$ in Table~3 and Fig.~1). Such process has been
invoked in numerous investigations dealing with open questions in CV
evolution (e.g. Patterson 1998, Kolb \& Baraffe 1999, Schwope et al
2002). (ii) In the second scenario, only gravitational radiation is
effective, but the secondary has lost thermal equilibrium after the
sudden turn-on of mass transfer causing its effective mass radius
exponent $\zeta$ to drop to a value near the adiabatic one,
$\zeta_{\rm ad} \simeq -1/3$, (model g\,+\,s with an arbitrarily
chosen $\zeta \simeq -0.2$). In this case, the orbital period would
currently be stagnating or increasing, $\dot M_1$ would exceed the
equilibrium value by about a factor of three, and the object would be
located near the ``tip of the flagpole'', the flag being the loop
described by the evolving CV in the $\dot M_1,P_{\rm orb}$-plane
(d'Antona et al. 1988, Kolb \& Baraffe 1999). It is noteworthy that
turn-on would take a couple of million years which agrees with the time
span over which the current spin-up rate in EX~Hya might have been
sustained if it started from a synchronized white dwarf. We can not
presently distinguish between the scenarios (i) and (ii). The latter
possibility might hold for EX~Hya as a single object, but not for
short-period CVs in general. A more precise primary mass would allow
to place tighter limits on the evolutionary state of EX~Hya.

\section{Conclusion}

Our high precision astrometric parallax has led to an internally
consistent picture of EX Hya as an intermediate polar. With a distance
of $d = 64.5\pm 1.2$\,pc, the luminosity-derived, time-averaged
accretion rate of the white dwarf in EX~Hya is $\dot M_{\rm
1,av}=(6.5\pm 1.5)\times 10^{-11}(M_1/0.5\,{\rm M}_\odot)^{-1.61}\,
{\rm M}_{\odot}{\rm yr}^{-1}$, valid over about the last 30 years. The
largest uncertainty in pinning down $\dot M_{\rm 1,av}$ is still the
poorly known primary mass $M_1$. The kinematic solution, the
interpretation of X-ray observations (given a somewhat lower weight),
and a range of other indirect arguments lead to a permitted range of
$M_1=0.40-0.70$\,\msun, with a preference for masses around
0.5\,\msun. Only for $M_1=0.6-0.7$\,\msun, would the time-averaged
accretion rate $\dot M_{\rm 1,av}=4\times 10^{-11}$\,\smyr\ coincide
with that expected from quasi-equilibrium mass transfer driven by
gravitational radiation plus the transfer of orbital into spin angular
momentum, fixed by the observed spin-up time scale of $(3.8\pm 0.2)\times
10^6$\,yrs (Hellier \& Sproats 1992). If, on the other hand,
$M_1=0.4-0.5$\,\msun, the implied higher accretion rate $\dot M_{\rm
1,av}=8\times 10^{-11}$\,\smyr\ requires either an additional angular
momentum loss process or the loss of thermal equilibrium of the
secondary star. In any case, a more precise value of $M_1$ would be an
invaluable ingredient for a more detailed discussion of the
evolutionary status of EX~Hya.

Spectroscopically, the secondary looks like a main sequence star as
expected in a CV with 98\,min orbital period ($M_2 \simeq
0.13$\,\msun), the permitted range of $M_2=0.07-0.10$\,\msun\
indicates, however, that it is undermassive, overluminous, and
expanded over a main ZAMS object of the same mass.

Our results confirm some of the basic concepts developed for CVs
over the last decades and demonstrates that accurate measurements of
parallax and velocity amplitudes can provide detailed information on
the evolutionary status of individual CVs.

\begin{acknowledgements} 
This research was supported in Germany by DLR/BMFT grant
50\,OR\,99\,03\,1.  In the Unites States, partial support for TEH,
BEM, and GFB for proposal \#9230 was provided by NASA through a grant
from the Space Telescope Science Institute, which is operated by the
Association of Universities for Research in Astronomy, Inc.,
\mbox{under NASA contract NAS} 5-26555. In the UK, BTG was supported
by a PPARC Advanced Fellowship. This research has made use of the
NASA/ IPAC Infrared Science Archive, which is operated by the Jet
Propulsion Laboratory, California Institute of Technology, under
contract with the National Aeronautics and Space Administration. This
publication also makes use of data products from the Two Micron All
Sky Survey, which is a joint project of the University of
Massachusetts and the Infrared Processing and Analysis
Center/California Institute of Technology, funded by the National
Aeronautics and Space Administration and the National Science
Foundation. We also would like to thank B. Skiff for pointing us to
the UCAC catalogue.
\end{acknowledgements}

\end{document}